\documentclass[10pt,twocolumn,tightenlines,aps,amsmath,floatfix,notitlepage,nofootinbib,showpacs,superscriptaddress]{revtex4-1}
\usepackage[T1]{fontenc}        
\usepackage{palatino}           
\usepackage{amsmath}            
\usepackage{amssymb}            
\usepackage{graphicx}           
\usepackage{rotating}
\usepackage{booktabs}
\usepackage{multirow}
\usepackage{pslatex}
\usepackage{epsfig}
\usepackage{epsf}
\usepackage[usenames]{color}
\usepackage[normalem]{ulem}
\usepackage{mathrsfs}           

\usepackage{caption}
\usepackage{hyperref}
\usepackage{siunitx}
\newcommand{\rem}[1]{}          
\newcommand{\nsf}[1]{I_{#1}^{\mathrm{NSF}}}
\newcommand{\ssf}[1]{I_{#1}^{\mathrm{SF}}}
\newcommand{\fft}{\frac{5}{2}}
\newcommand{\bff}{\bar{\frac{1}{2}}}
\newcommand{\mcn}[3]{\multicolumn{#1}{#2}{#3}}

\begin{document}

\title{Experimental determination of the magnetic interactions of frustrated Cairo pentagon lattice materials}
\author{Manh Duc Le}
\email{duc.le@stfc.ac.uk}
\affiliation{ISIS Neutron and Muon Source, Rutherford Appleton Laboratory, Chilton, Didcot, OX11 0QX, UK}
\affiliation{IBS Research Center for Correlated Electron Systems, Seoul National University, Seoul 08826, Korea}
\author{Elisa M. Wheeler}
\affiliation{Institut Laue-Langevin, Avenue des Martyrs, CS 20156, F-38042 Grenoble C\'edex 9, France}
\affiliation{Cellule DATA, Universit\'e de Strasbourg, 20a rue Ren\'e Descartes, F-67000 Strasbourg, France}
\author{Jaehong Jeong}
\affiliation{IBS Research Center for Correlated Electron Systems, Seoul National University, Seoul 08826, Korea}
\affiliation{Department of Physics and Astronomy, Seoul National University, Seoul 08826, Korea}
\author{K. Ramesh Kumar}
\affiliation{IBS Research Center for Correlated Electron Systems, Seoul National University, Seoul 08826, Korea}
\author{Seongsu Lee}
\affiliation{Neutron Science Division, Korea Atomic Energy Research Institute, Daejeon 34057, Korea}
\author{Chang-Hee Lee}
\affiliation{Neutron Science Division, Korea Atomic Energy Research Institute, Daejeon 34057, Korea}
\author{Myeong Jun Oh}
\affiliation{Department of Physics, Kyungpook National University, Daegu 41566, Korea}
\author{Youn-Jung Jo}
\affiliation{Department of Physics, Kyungpook National University, Daegu 41566, Korea}
\author{Akihiro Kondo}
\affiliation{Institute for Solid State Physics, The University of Tokyo, Kashiwa, Chiba 277-8581, Japan}
\author{Koichi Kindo}
\affiliation{Institute for Solid State Physics, The University of Tokyo, Kashiwa, Chiba 277-8581, Japan}
\author{U. Stuhr}
\affiliation{Laboratory for Neutron Scattering, Paul-Scherrer-Institut, 5232 Villigen PSI, Switzerland}
\author{B. F{\aa}k}
\author{M. Enderle}
\affiliation{Institut Laue-Langevin, Avenue des Martyrs, CS 20156, F-38042 Grenoble C\'edex 9, France}
\author{Dmitry Batuk}
\affiliation{EMAT, University of Antwerp, Groenenborgerlaan 171, B-2020 Antwerp, Belgium}
\author{Artem M. Abakumov}
\affiliation{Skolkovo Institute of Science and Technology, Nobel str. 3, 143026 Moscow, Russia}
\author{Alexander A. Tsirlin}
\affiliation{Experimental Physics VI, Center for Electronic Correlations and Magnetism, University of Augsburg, 86159 Augsburg, Germany}
\author{Je-Geun Park}
\affiliation{IBS Research Center for Correlated Electron Systems, Seoul National University, Seoul 08826, Korea}
\affiliation{Department of Physics and Astronomy, Seoul National University, Seoul 08826, Korea}
\affiliation{Center for Quantum Materials, Seoul National University, Seoul 08826, Korea}

\date{\today}


\begin{abstract}
We present inelastic neutron scattering measurements of the Cairo pentagon lattice magnets Bi$_2$Fe$_4$O$_9$ and Bi$_4$Fe$_5$O$_{13}$F, supported by high field magnetisation measurements of Bi$_2$Fe$_4$O$_9$.
Using linear spin wave theory and mean field analyses we determine the spin exchange interactions and single-ion anisotropy in these materials.
The Cairo lattice is geometrically frustrated and consists of two inequivalent magnetic sites, both occupied by Fe$^{3+}$ ions and connected by two competing nearest neighbour interactions.
We found that one of these interactions, coupling nearest neighbour spins on the three-fold symmetric sites, is extremely strong and antiferromagnetic.
These strongly coupled dimers are then weakly coupled to a framework formed from spins occupying the other inequivalent site.
In addition we found that the Fe$^{3+}$ $S=5/2$ spins have a non-negligible single-ion anisotropy, 
which manifests as a spin anisotropy gap in the neutron spectrum and a spin-flop transition in high field magnetisation measurements.

\end{abstract}


\maketitle
%
%



\section{Introduction} \label{sec-intro}

\begin{figure}
  \begin{center}
    \includegraphics[width=0.95\columnwidth,viewport=12 16 580 461]{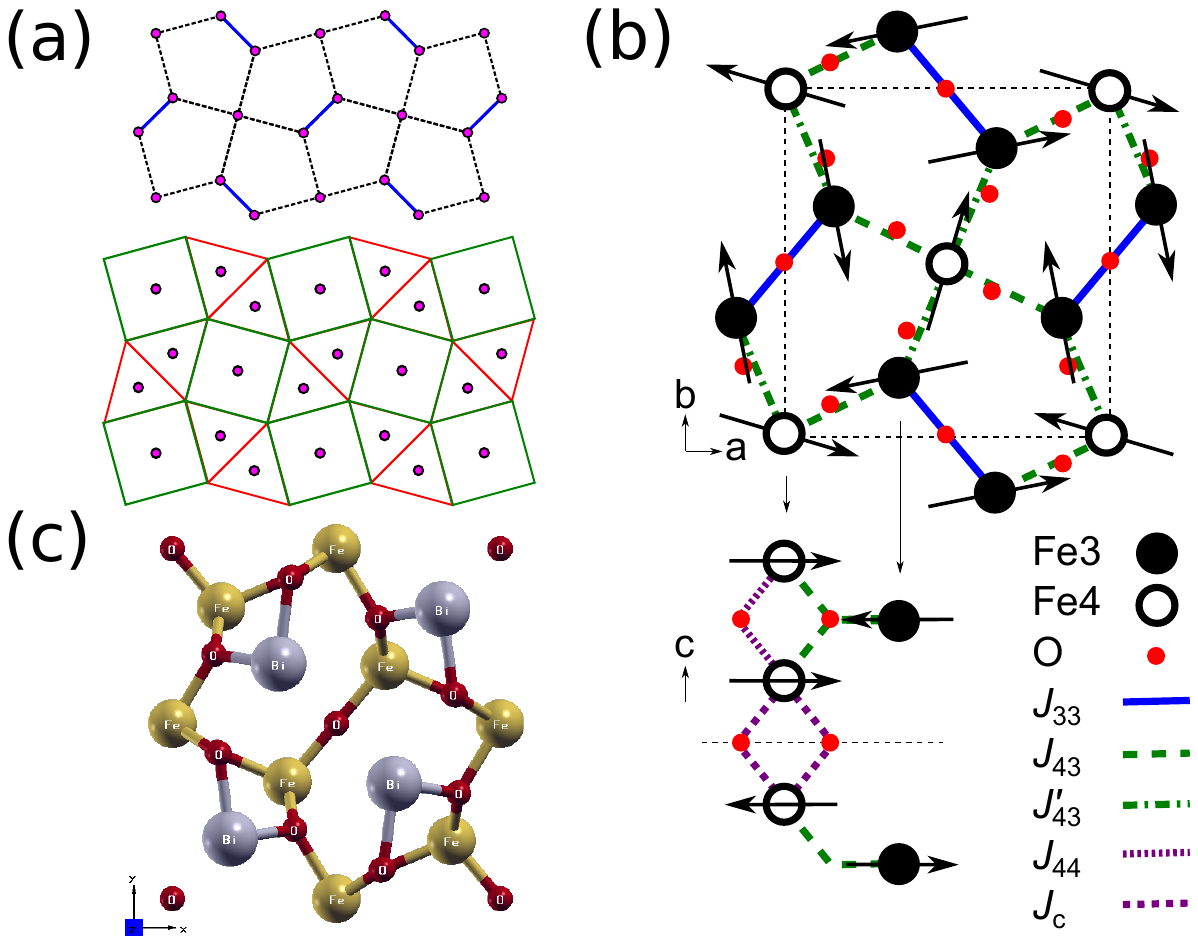}
    \caption{
(a) The Cairo lattice, with magnetic ions at the vertices of a lattice of irregular pentagons. 
Alternatively, the magnetic ions may be visualised as being enclosed by an edge-shared network of squares and equilateral triangles.
(b) Sketch of the crystal and magnetic structure of Bi$_2$Fe$_4$O$_9$ showing only Fe and O atoms and the Fe-O-Fe superexchange pathways.
The Fe$^3$ sites are the same as three-fold symmetric sites in the Cairo lattice, whilst pairs of Fe$^4$ spins occupy positions above and below the Cairo-lattice four-fold symmetric sites.
(c) View of the crystal structure perpendicular to the $c$-axis, with Bi atoms in grey, Fe in yellow, O in red.
    }
    \label{fg:struct}
  \end{center}
\end{figure}

Competing exchange interactions in geometrically frustrated magnets may lead to a large number of degenerate ground states~\cite{frustratedbook}.
The prototypical case is the Ising antiferromagnet on a triangular lattice, in which only half of the nearest neighbour interactions may be satisfied by any particular spin arrangements.
In the more conventional Heisenberg case, where the spins are free to adopt a range of orientations, a unique ground state with spins oriented at \ang{120} to their neighbours may be stabilised.
Besides triangles, other odd-vertex polygons will also exhibit geometrical frustration, however a tiling of these shapes in a regular crystal is forbidden by symmetry.
Nonetheless, lattices exists which consists of tilings of irregular odd-vertex polygons, of which the Cairo pentagonal lattice is one.
In this case, the pentagons have four equal and one unequal sides, as illustrated in Fig.~\ref{fg:struct}(a).

This gives rise to two inequivalent sites with three- and four-fold rotational symmetry perpendicular to the pentagon plane, which are connected by two competing nearest neighbour interactions.
One coupling is along the 4 equal edges which links three-fold to four-fold sites ($\mathcal{J}_{43}$), and the other goes via the single unequal edge of the pentagon, which links two three-fold sites ($\mathcal{J}_{33}$).
When these interactions are antiferromagnetic the system is geometrically frustrated and a frustration index may be constructed as $x=\mathcal{J}_{43}/\mathcal{J}_{33}$.
This is small when $\mathcal{J}_{33}$ dominates, leading the nearest-neighbor three-fold sites (denoted Fe$^3$), connected by $\mathcal{J}_{33}$, to form antiparallel pairs which may be thought of as dimers. 
Neighboring three-fold pairs are connected via the four-fold sites (denoted Fe$^4$) and the competing $\mathcal{J}_{43}$ interaction.
If $\mathcal{J}_{43}$ is small, then for classical (Heisenberg) spins, a non-collinear spin configuration analogous to the \ang{120} structure of the triangular lattice antiferromagnets is theoretically predicted~\cite{rousochatzakis2012}.
In this \emph{orthogonal} structure, shown in Fig.~\ref{fg:struct}(b), neighboring three-fold pairs are aligned at \ang{90} to each other, as are next-nearest neighbor four-fold spins.
For quantum spins, in the small $x$ limit, instead of the orthogonal structure, a collinear, partial ordered structure is predicted to be stabilised, where half of the three-fold sites do not order.
Finally, when $\mathcal{J}_{43}$ dominates, at large $x$, a collinear ferrimagnetic structure is predicted where the three-fold and four-fold sublattices are each individually ferromagnetic but are anti-aligned with each other.
In between these limits, when $x\approx\sqrt{2}$, \citet{rousochatzakis2012} predict that a spin nematic phase would be realised for quantum spin.

In the two experimental realisations of the Cairo lattice studied in this work, Bi$_2$Fe$_4$O$_9$ and Bi$_4$Fe$_5$O$_{13}$F,
the $\mathcal{J}_{33}$ interaction is expected to dominate due to the 180$^{\circ}$ Fe-O-Fe bond angle which is favourable for large antiferromagnetic superexchange interactions,
in contrast to the 120-130$^{\circ}$ bond angles of the other nearest neighbour interactions.
Indeed the orthogonal structure is observed at low temperatures in both Bi$_2$Fe$_4$O$_9$~\cite{ressouche2009} and Bi$_4$Fe$_5$O$_{13}$F~\cite{abakumov2013,tsirlin2017}. 
This orthogonal structure 
(expected in the case of 
low frustration index $x$) is non-collinear like the triangular \ang{120} structure,
and may lead to interesting magnetic excitations, as was recently observed for the \ang{120} system (Y,Lu)MnO$_3$~\cite{oh2013,oh_ymno3_magnonphonon}. 

In order to verify the exchange interactions, and thus experimentally determine $x$, and to compare with the theoretical phase diagram of~\cite{rousochatzakis2012}, we have measured the magnon spectrum of Bi$_2$Fe$_4$O$_9$, and the related material Bi$_4$Fe$_5$O$_{13}$F, which is formed by interleaving additional layers of Fe$^{3+}$ ($S=5/2$) spins (denoted Fe$^{\mathrm{int}}$) in between the 4-fold sites along $c$ between the pentagonal layers.

\section{Methods} \label{sec-exp}

Inelastic neutron scattering measurements were performed on single-crystals of Bi$_2$Fe$_4$O$_9$ on the triple-axis spectrometers EIGER (at the Paul-Scherrer-Institut, Switzerland)~\cite{eiger} and IN20 (at the Institut-Laue-Langevin, France)~\cite{in20}.
A co-aligned array of 6 crystals was used for the EIGER measurements with total sample mass $\approx$0.6~g, whilst the IN20 measurements used only the largest crystal with mass 0.28~g.
The crystals were grown in Bi$_2$O$_3$ flux in Pt crucibles, and their crystallinity and orientations were checked using the Bio-D diffractometer at HANARO, Korea~\cite{bio-d_paper}.
A small 15~mg single crystal was used for high field magnetisation measurements at ISSP, Japan, in pulsed fields up to 65~T, with the crystals oriented with the field along $(110)$ and $(001)$ in turn.

In the EIGER measurements the crystals were mounted with $[hh0]-[00l]$ directions in the horizontal scattering plane, whilst on IN20 the $[h00]-[0k0]$ directions were horizontal.
In the IN20 measurements, small out-of-plane momentum transfers were accessible by tilting the cryostat.
Both spectrometers were operated with fixed final energy $E_F=14.7$~meV in the $W$-configuration with open collimation.
We used the double-focussing Si(111) monochromator and PG(002) analyser on IN20, whilst a focused PG(002) monochromator and analyser set up was used on EIGER.
Both set-ups used a PG-filter before the analyser.
Because of the lower flux of EIGER, measurements there were carried out at 200~K, just below $T_N$=245~K, where the larger magnon population factor enhances the signal.
In addition to unpolarised measurements~\cite{ill_4-01-1428} of the magnon dispersion at 2~K on IN20, we also measured the $XYZ$ polarisation dependence at selected Brillouin zone center and zone edge positions.
In this case, the horizontally focussed Heusler monochromator and analyser was used with a flipper on the scattered beam and a PG filter in front of the analyser.
Helmholtz coils were used to align the neutron polarisation to be parallel to the momentum transfer ($x$), perpendicular to this in the horizontal scattering plane ($y$) and vertically ($z$).

Supplementary inelastic neutron scattering measurements~\cite{rb1820598} on polycrystalline powders of Bi$_2$Fe$_4$O$_9$ were taken using the Merlin~\cite{guidi_merlin} and MAPS~\cite{maps} time-of-flight (ToF) spectrometers at the ISIS facility, UK.
The 20~g powder sample was synthesised from a solid state reaction in air and checked to be single phased using a Rigaku MiniFlex X-ray diffractometer.
Merlin was operated in rep-rate multiplication mode with the gadolinium Fermi chopper running at 600~Hz and focused incident energies of 120~meV and 180~meV.
On MAPS, we used $E_i$=300~meV with the "sloppy" chopper running at 600~Hz.
Finally, inelastic neutron scattering measurements~\cite{ill_4-01-1427} on polycrystalline powders of Bi$_4$Fe$_5$O$_{13}$F were performed on the IN4~\cite{in4c} ToF spectrometer at ILL, France.
The 6.4~g powder sample was synthesised by a solid state reaction in evacuated sealed quartz tubes and used previously for diffraction measurements~\cite{abakumov2013, tsirlin2017}.
We used incident neutron energies of 150, 66, and 16.6~meV with the Fermi chopper running at 333 Hz for 150 and 66 meV and 200 Hz for 16.6 meV.
Additional measurements at higher energies on the same sample was carried out using the MARI spectrometer at ISIS.
The IN4 data was processed using the LAMP~\cite{lamp} software with subsequent checks using Mantid~\cite{mantid}.
MAPS, Merlin and MARI data were processed using Mantid.
We used the SpinW~\cite{toth_spinw, spinwwebsite} and McPhase~\cite{mcphasedmd, mcphasewebsite} software packages to model the data.

\section{Inelastic Neutron Scattering} \label{sec-ins}

\subsection{Bi$_2$Fe$_4$O$_9$} \label{sec-ins-bfo}

\begin{figure*}
  \begin{center}
    \includegraphics[width=0.90\textwidth,viewport=17 17 824 578]{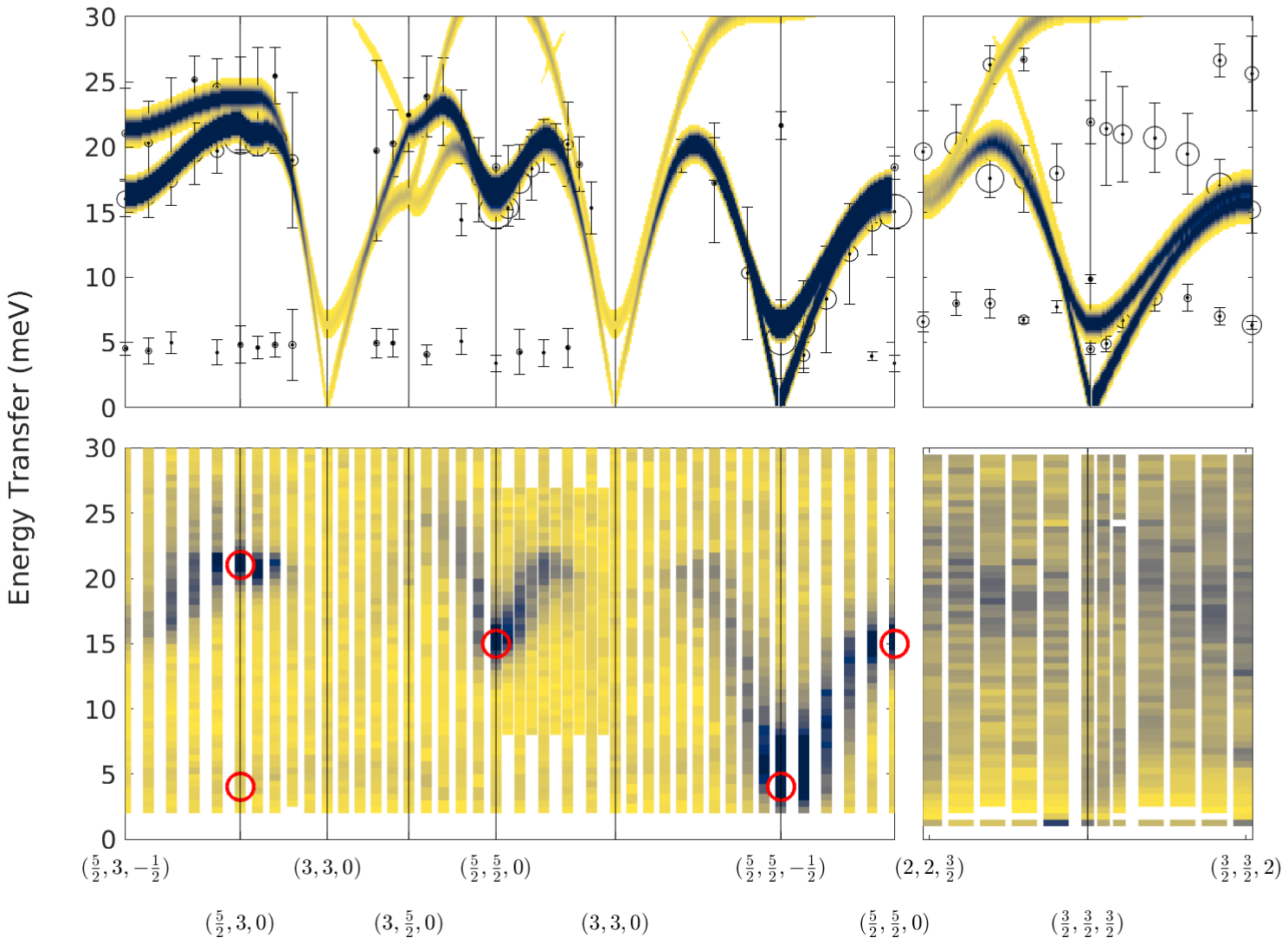}
    \caption{
The magnon dispersion of Bi$_2$Fe$_4$O$_9$ at 2~K measured by inelastic neutron scattering on IN20 (bottom left),
and calculated from the spin-wave model described in the text (top left), together with that measured at 200~K on EIGER (right panels).
Markers in the upper panels indicated the fitted peak positions with the marker size proportional to the fitted peak areas, and the errorbars proportional to the fitted widths.
The colormap shows the measured $\chi''(\mathbf{Q},\omega)$ on an arbitrary scale.
The flat mode below 10~meV and that around 20~meV along $[00L]$ in the Eiger data are believed to be phonons, whilst the other modes are well fitted by the spin wave model.
Large red circles indicate the points where polarised neutron measurements were made.
    }
    \label{fg:in20}
  \end{center}
\end{figure*}

Figure~\ref{fg:in20} shows the single-crystal unpolarised data for Bi$_2$Fe$_4$O$_9$ collected on IN20 (at 2~K) and EIGER (at 200~K).
In order to compare the two different temperatures, the measured scattering intensity $S(\mathbf{Q},\omega)$ was converted to the dynamical susceptibility $\chi''(\mathbf{Q},\omega)$
via the fluctuation-dissipation theorem, $\chi''(\mathbf{Q},\omega) \propto (1-\exp(-\beta\hbar\omega)) S(\mathbf{Q},\omega)$ and it is $\chi''(\mathbf{Q},\omega)$ which is plotted on an arbitrary scale.
The measured dispersion at low temperature is very similar to a recent single crystal inelastic neutron scattering study~\cite{beauvois2019}.
Through most of the Brillouin zone, only a single intense dispersive mode was observed in the energy transfer range up to 30~meV.
A much weaker, almost flat mode at around 5~meV energy transfer was also seen on both IN20 and Eiger, and
another faint dispersive mode between 15 and 25~meV in the $(00L)$ direction is seen in the Eiger data at 200~K but not in the IN20 data.
These modes are not well described by linear spin wave theory (see section~\ref{sec-ins-swt}) and could be phonons, whilst 
the higher energy mode
may arise from a spurious signal due to higher order scattering from the monochromator and analyser.
This "2$k_i$=3$k_f$ spurion" comes from the second order scattering of neutrons from the monochromator which matches the energy for third order scattering from the analyser
and appears as an apparent inelastic band at an energy transfer of $\frac{5}{4}$ of $E_f$, which in our case corresponds to 18.4~meV.
This only applies to the EIGER experiments where PG(002) was used for both the analyser and monochromator.
For the IN20 measurements, a Si(111) monochromator combined with a PG(002) analyser was used and the different lattice parameters means the different higher orders do not coincide in energy.

Table~\ref{tab:lpa} summarises the longitudinal polarisation analysis measurements on IN20 at 2~K at selected reciprocal lattice positions $(\mathbf{Q}, \omega)$.
The spectrometer was set to these positions and the cross-sections with the neutron beam polarised along the directions $x||\mathbf{Q}$, $y\perp \mathbf{Q}$ in the horizontal plane and $z$ vertical
was measured in the spin-flip (SF) and non-spin-flip (NSF) channels.
We did not perform a rocking scan but left the spectrometer at the fixed $(\mathbf{Q}, \omega)$ for the six measurements per position shown.
From these cross-sections, signals proportional to the spin fluctuations in the $y$ and $z$ directions may be determined using the equations~\cite{enderle2014}

\begin{eqnarray} \label{eq:pol} \nonumber
M_{yy} = \langle M_{\perp y} M_{\perp y}^{\dagger} \rangle_{\omega} &=& \sigma_y^{\mathrm{NSF}} - \sigma_x^{\mathrm{NSF}} \\
                                                                    &=& \sigma_x^{\mathrm{SF}} - \sigma_y^{\mathrm{SF}}   \\ \nonumber
M_{zz} = \langle M_{\perp z} M_{\perp z}^{\dagger} \rangle_{\omega} &=& \sigma_z^{\mathrm{NSF}} - \sigma_x^{\mathrm{NSF}} \\ \nonumber
                                                                    &=& \sigma_x^{\mathrm{SF}} - \sigma_z^{\mathrm{SF}}
\end{eqnarray}

\noindent where $M_{\perp}$ is the component of the magnetic scattering vector perpendicular to $\mathbf{Q}$
and $\langle \ldots \rangle_{\omega}$ indicates the time Fourier-transformed thermal average.
We have also assumed that the chiral ($\propto \langle \mathbf{M} \times \mathbf{M} \rangle_{\omega}$) and nuclear-magnetic interference 
($\propto \langle N \mathbf{M} \rangle_{\omega}$, where $N$ is the nuclear scattering amplitude)
terms in the cross-section are negligible compared to the strong magnetic signal.
This means that the incident polarisation was always kept the same and only the flipper on the scattered beam was turned on (for SF) or off (for NSF).
The measurements in table~\ref{tab:lpa} show that the low energy flat mode around 4-5~meV is non-magnetic,
as indicated by the $(\frac{5}{2}30)$ 4~meV point,
implying that this mode is most likely a phonon. 
The polarised measurements also show that the magnetic fluctuations in the zone centre and zone boundaries are almost entirely out-of-plane (along $z||c$),
consistent with an easy plane single-ion anisotropy as will be discussed in section~\ref{sec-ins-swt}.

\begin{table*} \renewcommand{\arraystretch}{1.3}
\begin{center}
  \begin{tabular}{@{\extracolsep{\fill}}r|rr|rr|rr||rr|rr}
  \hline
      $(hkl), \hbar\omega$ &$\nsf{x}$ &$\ssf{x}$&$\nsf{y}$ &$\ssf{y}$&$\nsf{z}$ &$\ssf{z}$ &  \mcn{2}{c|}{$M_{yy}$}  & \mcn{2}{c}{$M_{zz}$}    \\
                           &          &         &          &         &          &          &     NSF     &     SF    &      NSF    &     SF    \\
  \hline
      $(\fft30)$, 4 meV    &  141     &    87   &   142    &   94    &   146    &   93     &   1(17)     &  -7(13)   &   5(17)     &  -6(13)   \\
      $(\fft30)$, 21 meV   &   41     &   388   &    46    &  347    &   335    &   71     &   5(9)      &  41(27)   & 294(19)     & 317(21)   \\
    $(\fft\fft0)$, 15 meV  &   56     &   499   &    73    &  533    &   539    &   71     &  17(11)     & -34(32)   & 483(24)     & 428(24)   \\
  $(\fft\fft\bff)$, 4 meV  &   42     &   213   &    43    &  256    &   256    &   33     &   1(9)      &  11(20)   & 214(17)     & 180(16)   \\
  \hline
  \end{tabular}
  \caption{
Longitudinal polarisation analysis measurements at selected zone centre and zone boundary positions. All values are counts per $4\times 10^6$ monitor counts (approximately 10 minutes),
except for the measurement at $(\fft30)$, 4meV which is in counts per $2\times10^7$ monitor counts.
Numbers in parenthesis are standard errors.
}
  \label{tab:lpa}
\end{center}
\end{table*}

In addition to the data below 30~meV shown in figure~\ref{fg:in20}, 
we had also measured peaks at around 35~meV which appeared dispersive in the IN20 data. 
However, later analysis showed that this was an experimental artifact, arising from the inelastic incoherent scattering of a sample Bragg peak from the analyser.
Preliminary spin wave theory calculations based on exchange parameters previously calculated by density functional theory (DFT)~\cite{pchelkina2013,tsirlin2017} showed two bands of excitations,
with the lower energy band associated with the interlayer exchange interaction and the upper coming from the in-plane interactions.
As we had apparently seen both bands in scans on IN20 up to 40~meV, we did not extend the measurements to higher energies at the time.
When subsequent analysis indicated that the upper mode is actually spurious scattering, we performed additional high energy measurements on powder samples.

\begin{figure}
  \begin{center}
    \includegraphics[width=1.00\columnwidth,viewport=0 0 461 344]{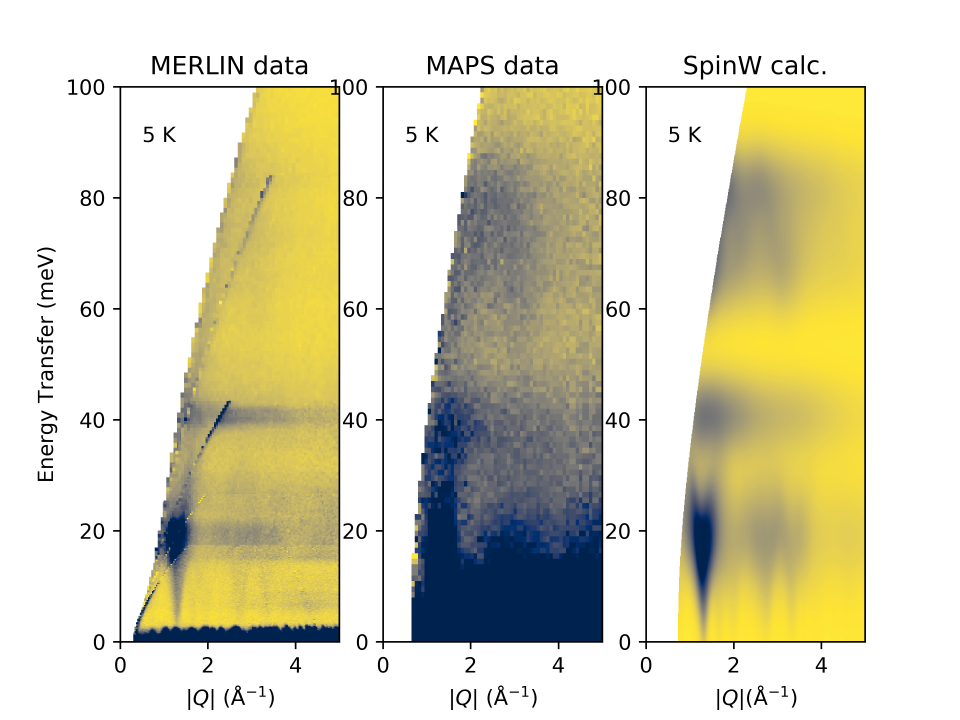}
    \caption{
Powder inelastic neutron scattering data on Bi$_2$Fe$_4$O$_9$ from MERLIN and MAPS at 5~K.
Data from multiple MERLIN runs with $E_i$=25, 38, 62, 120 and 180~meV is shown overlapped on the left panel. MAPS data shows the $E_i$=300~meV run.
Several magnon bands up to 90~meV are visible as described in the text.
    }
    \label{fg:bfo_slices}
  \end{center}
\end{figure}

\begin{figure}
  \begin{center}
    \includegraphics[width=0.85\columnwidth,viewport=42 7 424 306]{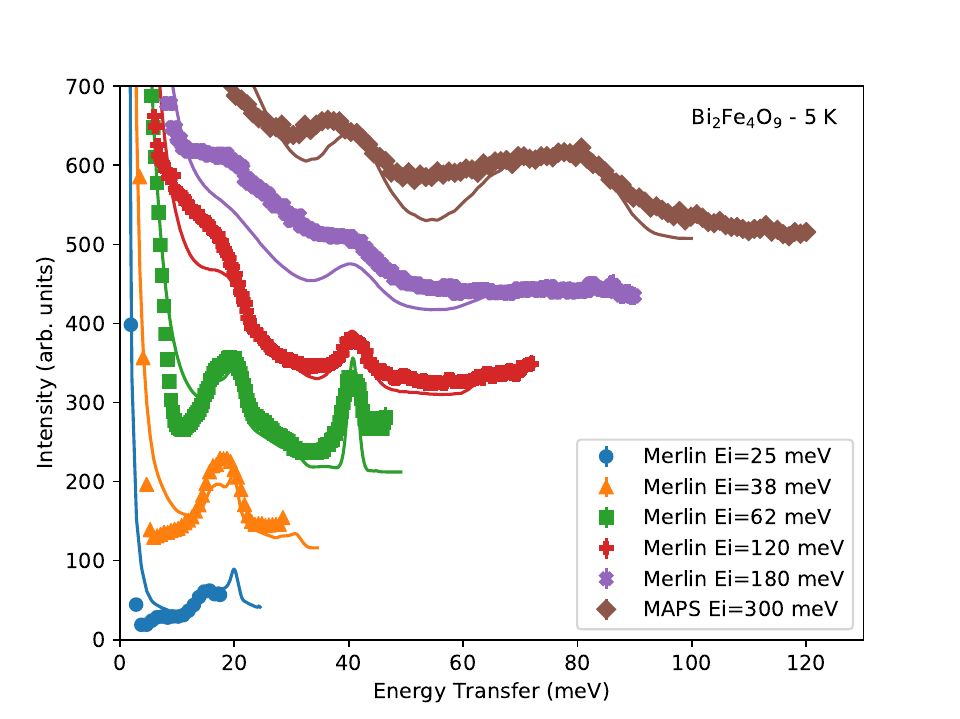}
    \caption{
Cuts along energy transfer of the powder inelastic neutron scattering data on Bi$_2$Fe$_4$O$_9$ from MERLIN and MAPS at 5~K, integrating over the low $|Q|$ region up to 3~\AA$^{-1}$. Solid lines are fits to a spin wave model described in the text.
    }
    \label{fg:bfo_cuts}
  \end{center}
\end{figure}

Figures~\ref{fg:bfo_slices}~and~\ref{fg:bfo_cuts} show the data on Bi$_2$Fe$_4$O$_9$ powder measured on MAPS and Merlin.
Additional scattering which follows the magnetic form factor was observed at $\approx$40~meV and between 60 and 80~meV.
Modelling the data, we found that the relatively narrow flat mode at 40~meV is associated with the in-phase precession of the ferromagnetically coupled spins which form the four-fold Fe$^4$ site in the Cairo lattice.
The energy of this mode thus serves to pin down the $J_{44}$ exchange parameter which couples these spins.
The in-plane dimer interaction, $J_{33}$, on the other hand determines the energy separation between the lower and upper magnon bands.
That we observed the upper band at relatively high energies (between 60 and 80~meV) thus implies a large |$J_{33}$| which is significantly larger than anticipated by DFT calculations.
Finally, the bandwidths of the magnon bands are determined by the other in-plane interactions (both upper and lower bands) and the interlayer coupling $J_c$ (lower band only).

\subsection{Bi$_4$Fe$_5$O$_{13}$F} \label{sec-ins-bfof}

\begin{figure}
  \begin{center}
    \includegraphics[width=1.00\columnwidth,viewport=0 0 461 344]{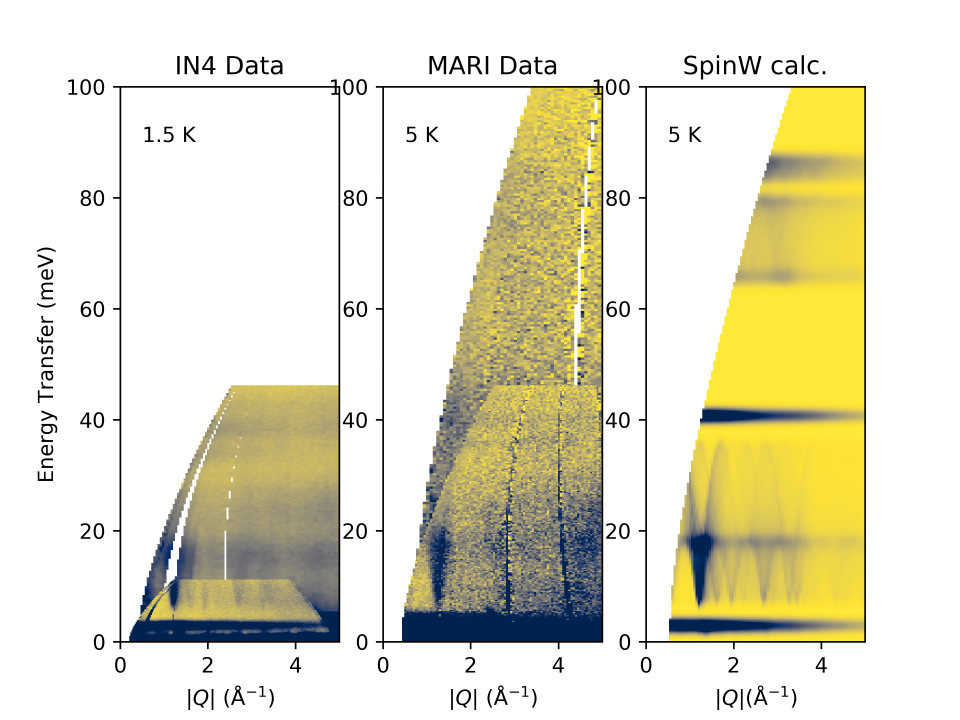}
    \caption{
Powder inelastic neutron scattering data on Bi$_4$Fe$_5$O$_{13}$F from IN4 and MARI at low temperatures.
Data from multiple runs with $E_i$=16.6 and 66~meV from IN4 and $E_i$=66 and 160~meV from MARI is shown overlapped.
Several magnon bands up to 90~meV similar to Bi$_2$Fe$_4$O$_9$ are observed as described in the text.
    }
    \label{fg:bfof_slices}
  \end{center}
\end{figure}

\begin{figure}
  \begin{center}
    \includegraphics[width=0.85\columnwidth,viewport=32 7 424 312]{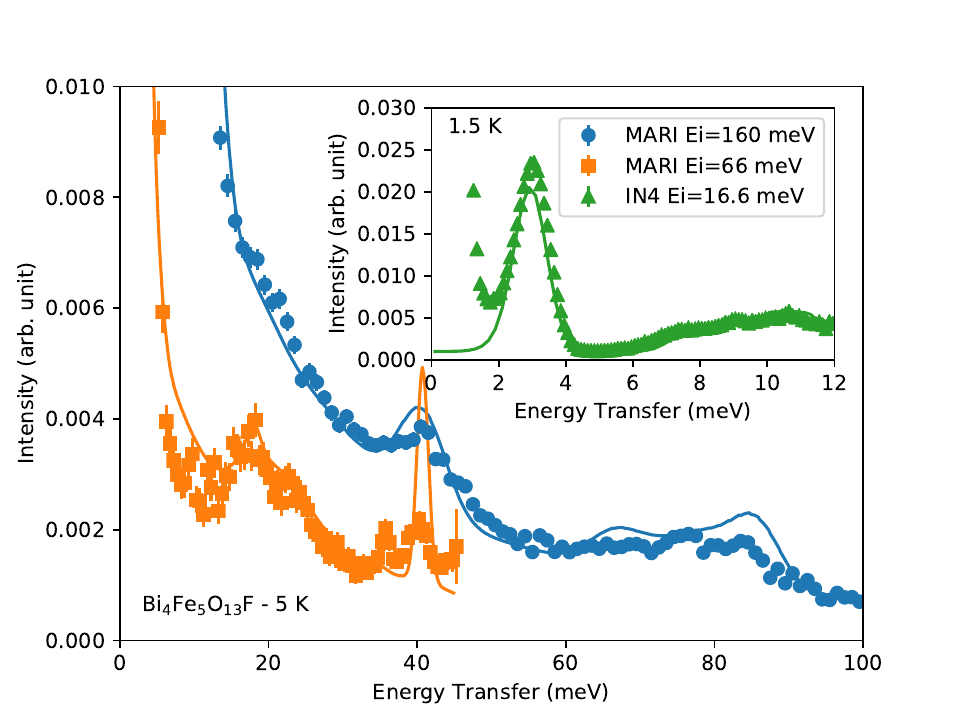}
    \caption{
Cuts along energy transfer of the powder inelastic neutron scattering data on Bi$_4$Fe$_5$O$_{13}$F from IN4 and MARI at low temperatures, integrating over a narrow region around the first magnetic Brillouin zone centre. Solid lines are spin wave model fits.
    }
    \label{fg:bfof_cuts}
  \end{center}
\end{figure}

\begin{figure}
  \begin{center}
    \includegraphics[width=0.85\columnwidth,viewport=37 7 424 308]{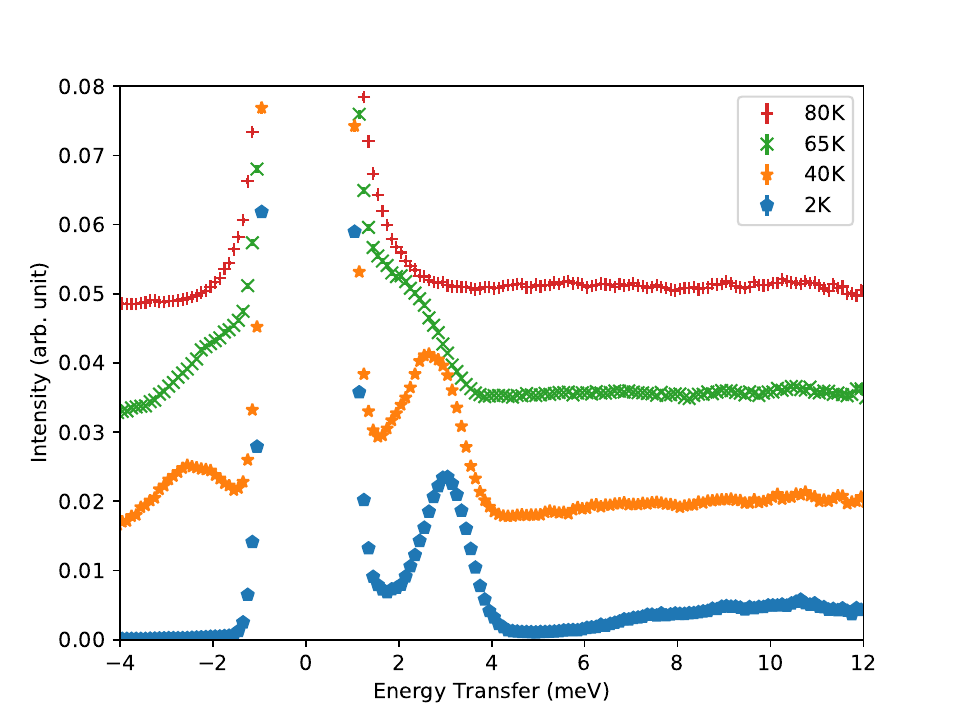}
    \caption{
Temperature dependence of the scattering around the first Brillouin zone centre as a function of temperature, as measured on IN4 with $E_i$=16.6~meV.
The mode at $\approx$3~meV is associated with the intermediate layer spins which become disordered above $T_1$=62~K:
it is clearly present in the low temperature phase below $T_1$, becomes quasi-elastic in the intermediate phase between $T_1$ and $T_2=71$~K, and disappears above $T_2$.
    }
    \label{fg:bfof_tdep}
  \end{center}
\end{figure}

Figures~\ref{fg:bfof_slices}~and~\ref{fg:bfof_cuts} show the inelastic neutron spectrum of \ powder Bi$_4$Fe$_5$O$_{13}$F as 2D contour maps and energy cuts respectively.
Magnetic excitations are seen up to $\approx$80~meV like in Bi$_2$Fe$_4$O$_9$.
Similarly, the pattern of a strongly dispersive band below 20~meV, a flat band around 40~meV and another band at higher energies is also seen.
The largest difference between the spectra of the two materials is an intense flat mode below the dispersive mode at $\approx$3~meV.
This low energy mode is observed only at low temperatures in phase I, and disappears above the $T_1$=62~K transition, as shown in figure~\ref{fg:bfof_tdep}.
It is thus associated with ordering of the intermediate Fe$^{\mathrm{int}}$ layer,
which is supported by spinwave modelling where the mode disappears if the moments on the Fe$^{\mathrm{int}}$ site is disordered.

In addition, the 40~meV flat band in Bi$_2$Fe$_4$O$_9$ appears to be split into two bands in Bi$_4$Fe$_5$O$_{13}$F.
Like in Bi$_2$Fe$_4$O$_9$, this mode is associated with the ferromagnetically coupled 4-fold symmetric sites Fe$^4$.
In addition these ions are coupled to the intermediate layer Fe$^{\mathrm{int}}$ sites by the interlayer $J_{c}$ interaction.
However, this antiferromagnetic interaction is not enough to lead to a splitting of the 40~meV band.
As explained in the next section, this splitting could be described in spin-wave theory only if the intermediate layer Fe$^{\mathrm{int}}$ spins are canted with respects to the Fe$^4$ spins.
Since neutron diffraction measurements show that this is not the case, we have been unable to model this splitting in the data.

\section{Spin Wave Theory} \label{sec-ins-swt}

\begin{figure}
  \begin{center}
    \includegraphics[width=0.85\columnwidth,viewport=23 372 580 825]{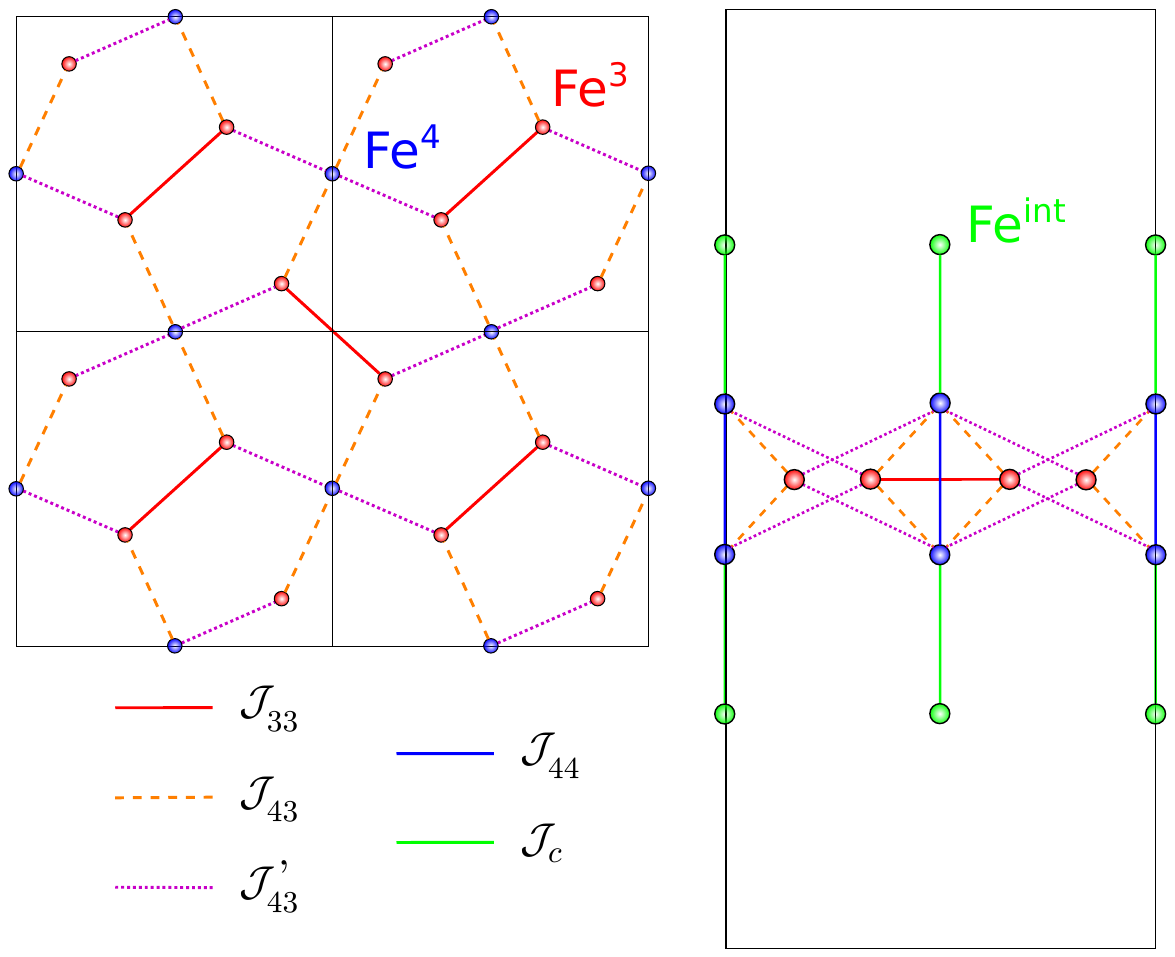}
    \caption{
Exchange couplings between Fe$^{3+}$ spins in the Cairo lattice compounds.
Blue circles indicate the 4-fold sites Fe$^4$, red circles are the 3-fold sites Fe$^3$ and green circles are the intermediate layer Fe$^{\mathrm{int}}$ sites in Bi$_4$Fe$_5$O$_{13}$F.
In Bi$_2$Fe$_4$O$_9$ there are no Fe$^{\mathrm{int}}$ sites, and instead the $J_c$ couplings link Fe$^4$ sites on adjacent pentagon layers.
    }
    \label{fg:couplings}
  \end{center}
\end{figure}

The exchange interaction topology for the two materials is shown in figure~\ref{fg:couplings}.
We found that a Heisenberg spin Hamiltonian with isotropic exchange interaction and single-ion anisotropy is enough to model the measured excitations,

\begin{equation} \label{eq-hamiltonian}
\mathcal{H} = \sum_{ij} J_{ij} \mathbf{S}_i\cdot\mathbf{S}_j + \sum_i A_i S_z^2(i)
\end{equation}

\noindent where positive $J>0$ indicates antiferromagnetic interactions and positive $A>0$ indicates an easy-plane single-ion anisotropy (SIA).

Following \emph{ab initio} calculations~\cite{tsirlin2017} which indicated that there are preferred easy-axes for each spin site within the $a-b$ plane,
we have also modelled the case where $A<0$ is negative where the spins prefer to lie parallel or antiparallel to an easy-axis direction along the bond between 3-fold and 4-fold sites.
However, the modelling shows that both axial and planar anisotropy scenarios can give almost the same dispersion and intensities 
(including polarised cross-sections at the measured zone centre and zone boundary points)
albeit with different values of $|A|$.

The exception to this is at low energies close to the zone centre or zone boundary, where in the easy-plane case the acoustic magnon is split into two modes.
One mode (involving precession out of the plane) is gapped, whilst the other modes (where the spins only precess within the plane) remains gapless.
In the easy-axis case, there is only a single gapped acoustic mode since any precession away from the preferred direction costs energy.
However, due to the presence of the magnetic Bragg peak and the incoherent elastic line it is hard to discern the presence of one or two modes.
The measured line widths of the constant-Q scans is quite broad and could be fitted by two peaks but there is no distinct minimum between them.
Polarised measurements at the zone boundary acoustic modes (e.g. $(330)$ 4~meV) would have been conclusive but the weak signal and lack of time meant we could not do this.
If the anisotropy is planar, this measurement would show that the magnetic fluctuations are polarised in the $a-b$ plane, 
whilst if the anisotropy is axial, the polarised measurement would show fluctuations to be mainly out-of-plane.

Since our data cannot distinguish between the easy-plane and easy-axis models,
we have chosen to use a spin-wave model with a single \emph{easy-plane} anisotropy term for Bi$_2$Fe$_4$O$_9$ for all sites for simplicity in the following analysis.
In the case of Bi$_4$Fe$_5$O$_{13}$F, the DFT calculations show that the magnitude of the SIA on the intermediate Fe$^{\mathrm{int}}$ sites is much larger than on the others.
We have thus modelled this with two easy-plane SIA terms, one for sites in the Cairo-lattice plane and one for the intermediate layer sites.

In terms of the exchange interactions, the real systems Bi$_2$Fe$_4$O$_9$ and Bi$_4$Fe$_5$O$_{13}$F deviate from the ideal Cairo pentagonal lattice in several respects.
First, the 4-fold site is occupied not by a single Fe$^{3+}$ spin, but rather by a pair of spins,
denoted Fe$^4$ in figure~\ref{fg:struct}(b) and~\ref{fg:couplings} offset above and below the pentagonal plane.
This leads additionally to an exchange interaction $J_{44}$ between these spins.
Second, due to an asymmetry in the positions of the oxygen ligands there are two different $\mathcal{J}_{43}$ interactions connecting the 4-fold sites and the 3-fold sites,
denoted $J_{43}$ and $J'_{43}$.
Finally there is a finite interlayer interaction $J_{c}$ through the 4-fold sites.
In Bi$_4$Fe$_5$O$_{13}$F this interaction connects the 4-fold sites above and below the pentagonal plane with an intermediate layer of Fe$^{3+}$ spins, denoted Fe$^{\mathrm{int}}$ in figure~\ref{fg:couplings}.
The 3-fold sites are denoted Fe$^3$.
We use this notation to ensure a uniform description of the different sites in the physical compounds and the theoretical Cairo lattice,
and to reconcile the different notations in the literature for Bi$_2$Fe$_4$O$_9$ and Bi$_4$Fe$_5$O$_{13}$F~\footnote{
In the literature for Bi$_2$Fe$_4$O$_9$, Fe$_1$ is used for the three-fold sites we label Fe$^3$ and Fe$_2$ for the sites we label Fe$^4$~\cite{ressouche2009,beauvois2019}.
In contrast, the literature for Bi$_4$Fe$_5$O$_{13}$F uses Fe$_1$ for the \emph{four-fold} site Fe$^4$, Fe$_2$ for the three-fold site Fe$^3$ and Fe$_3$ for the intermediate layer site Fe$^{\mathrm{int}}$~\cite{tsirlin2017}.}

\begin{table} \renewcommand{\arraystretch}{1.3}
\begin{center}
  \begin{tabular}{@{\extracolsep{\fill}}r|rr|r}
  \hline
                                      & \multicolumn{2}{c|}{Bi$_2$Fe$_4$O$_9$}                     &  Bi$_4$Fe$_5$O$_{13}$F \\
                                      & \qquad  This work        & \qquad Ref~\cite{beauvois2019}  &                        \\
  \hline
      $J_{33}$                        &    27.6(6)   \ \ \ \       &    24.0(8)                    &    29(2)  \ \ \ \ \    \\
      $J_{43}$                        &     3.1(2)   \ \ \ \       &     2.9(1)                    &     4.1(3)  \ \        \\
      $J'_{43}$                       &     6.5(2)   \ \ \ \       &     6.3(2)                    &     6.4(1)  \ \        \\
      $J_{44}$                        &    -0.22(3)  \ \           &     3.7(2)                    &    -0.3(1)  \ \        \\
      $J_{c}$                         &     1.39(5)  \ \           &     1.3(1)                    &     0.49(5)            \\
  \hline
      $A$                             &     0.096(5)               &                               &     0.04(2)            \\
      $A'$                            &                            &                               &     0.3(1)  \ \        \\
  \hline
  \end{tabular}
  \caption{
Spin wave exchange parameters fitted to data in meV. Positive values indicate antiferromagnetic exchange.
$A$ denotes the single-ion anisotropy term on the 3-fold and 4-fold sites of the Cairo plane, whilst $A'$ is the term which applies to the
intermediate layer spins in Bi$_4$Fe$_5$O$_{13}$F only.
}
  \label{tab:parameters}
\end{center}
\end{table}

As noted in the previous section, the in-plane Cairo dimer interaction $J_{33}$ splits the dispersive modes into two bands.
The upper band involves mostly precession of the spins on the 3-fold sites.
These spins are antiferromagnetically aligned, so that as $J_{33}$ becomes larger in the positive sense (becomes more antiferromagnetic),
this configuration is increasingly stabilised so that it costs more energy to cause the spins to precess away from rigid anti-alignment.
Our observation that the upper bands are between 60 and 80~meV thus points to a surprisingly large antiferromagnetic $J_{33}$, in agreement with~\cite{beauvois2019}.

The out-of-phase precession of the ferromagnetically aligned pair of spins above and below the Cairo 4-fold site yields a flat band at intermediate energy.
A larger ferromagnetic (more negative) $J_{44}$ exchange between these spins will cost this out-of-phase motion more energy
and hence push the mode up in energy, whilst more antiferromagnetic $J_{44}$ will push it down in energy.
We observed a flat mode at around 40~meV which points to a moderate \emph{ferromagnetic} $J_{44}$ in contrast with DFT calculations.
It is also counter to the findings of~\citet{beauvois2019} because they posit the flat band to be at $\approx$19~meV and so deduced a moderate \emph{antiferromagnetic} $J_{44}$.
However, 19 meV is very close to an energy transfer of 18.4 meV, which corresponds to a well-known (so-called "$2k_i=3k_f$") spurion of a triple-axis spectrometer
operating with a final wave vector of $k_f=2.662$~\AA$^{-1}$, as was the case in~\cite{beauvois2019}.
We note that we also observe scattering at around this energy in our EIGER data which does not fit the spin wave model and may also be due to this same spurion.

In Bi$_4$Fe$_5$O$_{13}$F at low temperatures, we apparently observe at least \emph{two} flat bands around 40~meV, as shown in figure~\ref{fg:bfof_slices} and~\ref{fg:bfof_cuts}.
If these bands are magnetic, they may only be explained in terms of spin wave theory by a canting of the intermediate site Fe$^{\mathrm{int}}$ spins with respects to the 4-fold site Fe$^4$ spins.
This would only arise if there is a very strong easy-axis single-ion anisotropy which would dominate over the exchange interaction $J_c$ between these spins which tends to keep them in (anti-)alignment.
We calculated that $A$ should be an order of magnitude larger than $J_c$ for this to pertain, which seems physically unlikely, given that the Fe$^{3+}$ ion has an $L=0$ ground state.
Moreover, neutron diffraction~\cite{tsirlin2017} measurements find that these spins (on sites Fe$^4$ and Fe$^{\mathrm{int}}$) are anti-parallel in Phase I at low temperature, with no canting.
There are also several optic phonon modes at these energies, but it is hard to discern if any of these modes are magnetic or phononic,
because all modes show an initial decrease in intensity from 1~\AA$^{-1}$ to 4~\AA$^{-1}$ followed by an increase in intensity above 6~\AA$^{-1}$.
Because of the unlikelihood of a large easy-axis single-ion anisotropy and since it would yield a magnetic structure contrary to that observed experimentally,
we have chosen the model with modest SIA which yields only a single flat mode.

Also in Bi$_4$Fe$_5$O$_{13}$F there is a low energy flat mode at $\approx$3~meV which is due to precession of the intermediate layer Fe$^{\mathrm{int}}$ spins.
The energy of this mode depends on $J_c$ and the SIA on the Fe$^{\mathrm{int}}$ spins $A'$, whilst $J_c$ also affects the lower magnon band between 6-25~meV.
We found that to fit both these features required a relatively large $A'$$=0.3$~meV compared to the value in Bi$_2$Fe$_4$O$_9$ of $A'$$=0.096$~meV,
which is itself surprising for Fe$^{3+}$ which has a $L=0$ ground state and is needed to explain the $\approx$5~meV anisotropy gap in Bi$_2$Fe$_4$O$_9$.
However, the large value of $A'$ compared to $A$ is consistent with DFT calculations of the SIA in Bi$_4$Fe$_5$O$_{13}$F~\cite{tsirlin2017}
which found that the magnitude of SIA on the Fe$^{\mathrm{int}}$ site is over 5 times that on the other sites.

Finally, the bandwidths of the lower and upper bands are determined by $J_{43}$ and $J_{43}'$, whilst the lower bands are additionally
modulated by $J_{c}$, since the upper bands involve mostly the 3-fold sites which are not coupled by $J_{c}$.

Taking these considerations into account together with the observed energies of the different magnon bands in the two compounds
we were able to obtain suitable starting parameters for a local minimum simplex search.
The fitted exchange interactions are given in table~\ref{tab:parameters}.
Neutron data files, spin wave models and data analysis scripts may be obtained from~\cite{cairogithub}.

\section{High field magnetisation} \label{sec-hf}

\begin{figure}
  \begin{center}
    \includegraphics[width=0.95\columnwidth,viewport=12 11 449 333]{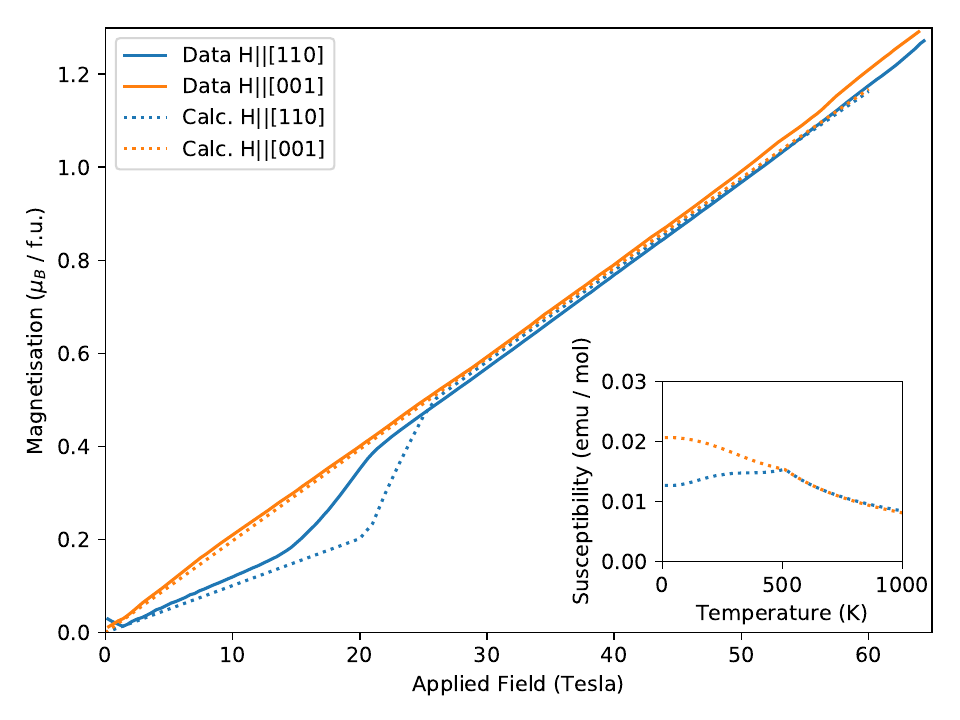}
    \caption{
High-field magnetisation of Bi$_2$Fe$_4$O$_9$ measured in a pulse field parallel to the $[110]$ and $[001]$ directions (solid lines)
and corresponding mean-field calculations (dotted lines). The calculated curves have been scaled to the measured ordered moment of 
Bi$_2$Fe$_4$O$_9$. The inset shows the calculated susceptibility at a field of 0.1~T.
A spin-flop transition is observed at $\approx$18~T in the data. In the model, this critical field depends on the single-ion anisotropy.
The SIA determined from a point charge model (see text) yields a critical field slightly higher than the observation but is qualitatively similar.
    }
    \label{fg:bfo_mag}
  \end{center}
\end{figure}

Figure~\ref{fg:bfo_mag} shows the high field magnetisation measured in a pulsed field and corresponding mean-field calculations.
A metamagnetic transition is observed around 18~T, and the mean-field calculations reveal it to be a spin-flop transition
where the spins cant out of the $a-b$ plane in order to minimise the Zeeman energy of the in-plane field.
The critical field depends primarily on the single-ion anisotropy which keeps the spins within the $a-b$ plane,
and to a lesser extent on the exchange interactions which prefers the orthogonal Cairo magnetic structure.

The mean field calculations used the exchange parameters fitted from the INS data as described above, but instead of a simple easy plane anisotropy term,
a point charge model was used to calculate the full crystalline electric field.
This is needed to use the \emph{intermediate coupling} scheme which includes the effect of the on-site Coulomb and spin-orbit interactions,
which we deemed to be more accurate than a simple spin-only approach.
We used the nominal charges ($+3|e|$ for Bi and Fe and $-2|e|$ for O) and calculated the effect of these charges on the magnetic $3d$ electrons of the Fe sites 
by including contributions from neighbouring charges up to 6.5~\AA~from a given Fe site, which covers up to second neighbouring Fe-ions.
These calculations found that the crystal field splitting on the 3-fold and 4-fold sites had similar magnitude, yielding an overall splitting of $\approx$0.4~meV.
This would correspond to a planar SIA term of $A$$_{\mathrm{pc}}=0.06$~meV, which is 60\% of the value found from fitting the spin anisotropy gap.
Scaling up the magnitude of the crystal field splitting to match the measure spin anisotropy gap would increase the spin-flop transition field by an equivalent amount.
Nonetheless, the chosen point charge model still over-estimates the critical field of the spin-flop transition, by $\approx$20\%.

In any case, the point charge calculations showed that additional anisotropy is present in the $ab$ plane, with the 3-fold sites prefering to align at 52$^{\circ}$ to the $a$-axis, and at 90$^{\circ}$, 180$^{\circ}$ and 270$^{\circ}$ to this angle.
The 4-fold sites, on the other hand preferred to align parallel and antiparallel to $c$ but are kept pinned to the $a-b$ plane by the exchange coupling with the 3-fold sites.
This is in contrast to the DFT calculations which found that all the moments preferred to lie within the $a-b$ plane.
We should note here that the point charge model is quite a coarse model and does not include many effects such as charge transfer and covalency which the DFT calculation does take into account. 

In addition, in order to match the measured slope of the magnetisation, we found that we needed to scale the mean-field calculations
(which assumed a full ordered moment of 5~$\mu_B$/Fe) by the ordered moment obtained by neutron diffraction measurements~\cite{ressouche2009}, $\approx$3.5~$\mu_B$.
The scaled magnetisation is shown in figure~\ref{fg:bfo_mag}.
We also calculated the temperature dependence of the magnetisation in a 0.1~T field, shown as an inset to figure~\ref{fg:bfo_mag}.
Because of the large $J_{33}$ interaction between pairs of 3-fold symmetric spins we have used a cluster mean-field method to calculate $M(T)$.
In the calculation, the pairs joined by $J_{33}$ are treated as a cluster using exact diagonalisation, but other interactions are treated in the mean-field approximation. 
This reduces the calculated $T_N$ compared with a pure mean-field calculation but still retains some effects of $J_{33}$.
The results agrees qualitatively with the measurements of~\citet{ressouche2009} but overestimates the ordering temperature by a factor of 2, yielding $T_N^{\mathrm{cmf}}=510$~K.
In contrast, a pure mean-field calculation gives $T_N^{\mathrm{cmf}}=1150$~K.

Finally we also applied the point charge calculations to Bi$_4$Fe$_5$O$_{13}$F, and found again that the 4-fold sites preferred to orient along $c$,
whilst the 3-fold sites preferred an in-plane direction offset 22$^{\circ}$ to the $a$-axis, and the intermediate layers to align along the $<$110$>$ directions.

\section{Discussions} \label{sec-conclusions}

In agreement with previous INS measurements~\cite{beauvois2019} we find a very strong dimer interaction $J_{33}$ between the 3-fold symmetric sites of the Cairo lattice
in both Bi$_2$Fe$_4$O$_9$ and Bi$_4$Fe$_5$O$_{13}$F, which puts these materials firmly within the \emph{orthogonal} phase of the theoretical phase diagram.
In both compounds the values of $J_{33}$ deduced from the spin wave spectra is almost three times that calculated from DFT~\cite{abakumov2013,tsirlin2017},
whilst deduced value of the other exchange interactions are of comparable magnitudes to the theoretical value.
The large $J_{33}$ also pushes the calculated mean-field ordering temperature to above 1000~K,
but goes some way to explaining the large Curie-Weiss temperature, $\theta_{\mathrm{CW}}\approx-1670~K$
(the sum of the deduced exchanges for Bi$_2$Fe$_4$O$_9$ is -1585~K).
We thus also used a cluster mean-field method that treats the dimer linked by $J_{33}$ as a cluster and applies exact diagonalisation to this cluster,
which is then coupled to the other spins in a mean-field fashion.
This calculation reduces the calculated $T_N^{\mathrm{cmf}}=510$~K which is still over twice the measured $T_N=245$~K.
Given the large $J_{33}$ one would expect the dimers to still be strongly coupled well above the overall ordering temperature $T_N$.
As noted by~\citet{beauvois2019}, this would lead to a correlated paramagnetic state which could give rise to novel phenomena.

The deduced exchange parameters for Bi$_2$Fe$_4$O$_9$ and Bi$_4$Fe$_5$O$_{13}$F are quite similar, with the main difference being the deduced interlayer-coupling,
which was found to be $J_c=1.39$~meV in Bi$_2$Fe$_4$O$_9$ and $J_c=0.49$~meV in Bi$_4$Fe$_5$O$_{13}$F.
This could perhaps explain part of the decrease in ordering temperature from $T_N=238$~K in Bi$_2$Fe$_4$O$_9$ to $T_N=178$~K in Bi$_4$Fe$_5$O$_{13}$F,
as the in-plane interactions in Bi$_4$Fe$_5$O$_{13}$F are deduced to be slightly larger than in Bi$_2$Fe$_4$O$_9$.
Bi$_4$Fe$_5$O$_{13}$F should thus also show the same correlated paramagnetic behaviour as Bi$_2$Fe$_4$O$_9$, but at a slightly lower temperature. 

Finally, in contrast to the expectation of the $L=0$ ground state of the Fe$^{3+}$ ion, we found that there is a non-negligible single-ion anisotropy in both Bi$_2$Fe$_4$O$_9$ and Bi$_4$Fe$_5$O$_{13}$F. 
This manifests in Bi$_2$Fe$_4$O$_9$ in a spin anisotropy gap of $\approx 5$~meV and a spin-flop transition at a critical field of $\approx 17$~T.
The gap is not directly measurable in Bi$_4$Fe$_5$O$_{13}$F because of the low energy flat mode at $\approx 3$~meV due to precession of intermediate layer spins in the lowest temperature ordered phase.
However, fitting the energy of this mode yields an even higher SIA than for Bi$_2$Fe$_4$O$_9$, which reflects the more distorted octahedral environment of the Fe$^{\mathrm{int}}$ spins,
and is in agreement with the larger SIA found for these sites by DFT calculations~\cite{tsirlin2017}.
This gives support to the model of~\citet{tsirlin2017} for the SIA being the origin of the series of phase transitions in Bi$_4$Fe$_5$O$_{13}$F.

In this model, the weak interlayer couplings means that the intermediate layer Fe$^{\mathrm{int}}$ spins are more susceptible to thermal fluctuations and become disordered first, at $T_1=62$~K.
Below this temperature, the large SIA on these Fe$^{\mathrm{int}}$ spins combined with the $J_c$ coupling constrains the Fe$^4$ spins to be antiparallel to the Fe$^{\mathrm{int}}$ spins.
The intra-layer interactions then favours a orthogonal magnetic structure with a particular mutual arrangement of adjacent Fe$^4$ spins described by a local vector chirality introduced in Ref.~\cite{tsirlin2017}.
However, the intrinsic SIA of the Fe$^3$ and Fe$^4$ sites prefers a different arrangement of the Fe$^4$ spins described by the opposite sign of the local vector chirality,
which is stabilised at high temperatures when the Fe$^{\mathrm{int}}$ spins become disordered, in the phase between $T_2=71$~K and $T_N$=178~K.
In between is a phase where the direction of preferred orientations of the Fe$^4$ spins determined by the competing SIA and intra-layer interactions
crosses-over giving a collinear, partial-ordered structure in the narrow temperature window between $T_1=62$~K and $T_2=71$~K.
Although this collinear structure is very similar to the theoretically predicted structure in the limit of small $\mathcal{J}_{43}/\mathcal{J}_{33}$~\cite{rousochatzakis2012}, it does not have the same origin. 
In the theory of~\citet{rousochatzakis2012}, which is based on a Heisenberg Hamiltonian without single-ion anisotropy, the collinear phase is driven by quantum fluctuations.
Whereas in the model of~\citet{tsirlin2017} it is a cross-over between two orthogonal ordered phases with opposite local vector chiralities, necessitating a collinear state in between.
Because we have a high spin ($S=5/2$) system, the threshold for the quantum fluctuations driven model is $\mathcal{J}_{43}/\mathcal{J}_{33} \lesssim 0.1$~\cite{tsirlin2017}.
Our larger value for $J_{33}$ than the DFT calculated values used by~\citet{tsirlin2017} suggests that we are approaching this limit in this case ($J_{43}/J_{33}=0.11$ and $J'_{43}/J_{33}=0.24$).
However, the experimental evidence for a relatively large single-ion anisotropy and the fact that the collinear phase exists only in a narrow range of temperature between two orthogonal phases
indicates that the single-ion anisotropy driven model for the origin of this phase is correct for Bi$_4$Fe$_5$O$_{13}$F.
This also suggests that there is scope for further theoretical work to extend the phase diagram to include a single-ion anisotropy term,
in order to better account for the behaviour of the materials which realise the Cairo lattice.

\section*{Acknowledgements}

The work at the IBS CCES was supported by the research program of Institute for Basic Science (IBS-R009-G1).
AT was supported by the Federal Ministry for Education and Research through the Sofja Kovalevskaya Award of Alexander von Humboldt Foundation.
Work of JGP was supported by the Leading Researcher Program of the National Research Foundation of Korea (Grant No. 2020R1A3B2079375).
Inelastic neutron scattering experiments were performed at the reactor of the Institute Laue-Langevin, Grenoble, France, at the Swiss spallation neutron source SINQ, at the Paul Scherrer Institute, Villigen, Switzerland, and at the ISIS Neutron and Muon Source, UK.
Experiments at ISIS were supported by a beamtime allocation RB1820598 from the Science and Technology Facilities Council.


\bibliographystyle{apsrev4-1}
\bibliography{mdlrefs}          


\end{document}